# Role of directional fidelity in multiple extreme performance of $F_1$-ATPase motor

Ruizheng Hou[1,3], Zhisong Wang[1,2,3*]

[1]Department of Physics, [2]NUS Graduate School for Integrative Sciences and Engineering,

[3]Center for Computational Science and Engineering, National University of Singapore,

Singapore 117542

*Correspondence author (phywangz@nus.edu.sg)

# ABSTRACT

Quantitative understanding of the best possible performance of nanomotors allowed by physical laws pertains to study of nanomotors from biology as well as nanotechnology. Biological nanomotor $F_1$-ATPase is the best available model system as it is the only nanomotor known for extreme energy conversion near the limit of energy conservation. Using a unified theoretical framework centred on a concept called directional fidelity, we analyze recent experiments in which $F_1$-motor's performance was measured for controlled chemical potentials, and expose from the experiments quantitative evidence for the motor's multiple extreme performance in directional fidelity, speed and catalytic capability close to physical limits. Specifically, the motor nearly exhausts available energy from the fuel to retain the highest possible directional fidelity for arbitrary load, encompassing the motor's extreme energy conversion and beyond. The theory-experiment comparison implies a tight chemomechanical coupling up to stalemate as futile steps occur but unlikely involve fuel consumption. The $F_1$-motor data also helps clarify relation between directional fidelity and experimentally measured stepping ratio.

## INTRODUCTION

Nanoscale motors are abundant[1-3] in biology, and are being actively pursued[4-10] in nanotechnology. But still inadequate are quantitative answers for a major scientific question, namely what is the best possible performance of nanomotors allowed by physical laws? Tackling such a challenging question needs theoretical as well as experimental inputs. A focus of previous studies is energy efficiency[11-15], perhaps motivated by the observation[16] that a biological nanorotor called $F_1$-ATPase or $F_1$-motor tends to convert almost all of the chemical energy of its fuel into work under a stalling load. As the only nanomotor known for extreme energy conversion near the limit of energy conservation, $F_1$-motor is the best available model system to study the question of best performance. Latest experiments[17] on this motor start to provide comprehensive performance data for precisely controlled chemical potentials and load. Applying a recently proposed concept[18, 19] to analyze the experiments, we show that the new data contain quantitative evidence for multiple extreme performance beyond yet encompassing the extreme energy conversion.

## METHODS

**Consensus chemomechanical cycle of $F_1$-motor.**

$F_1$-motor[1, 2] makes a 120-degree directional rotation[20] by hydrolyzing one fuel molecule called ATP (adenosine triphosphate) into ADP (adenosine diphosphate) and phosphate. The chemomechanical coupling between the motor's fuel consumption and rotation has been clarified by previous studies[1, 2, 21-23], and is summarized in Fig. 1. The motor has four major states[1, 2, 21] which each possess a unique coupling between the orientation of the rotor subunit and the chemical states of the three stator subunits responsible for catalyzing the fuel reaction. ATP hydrolysis (i.e., state 1 → state 2 in Fig. 1) and ATP binding (state 3 → state 4) occur when the rotor subunit has fixed orientations. Displacements of the rotor subunit occur upon release of fuel remnants: a 40º substep is triggered by phosphate dissociation (state 2 → state 3), followed by an 80º substep triggered by ADP dissociation (state 4 → state 1). A full 120º rotational step occurs by four sequential transitions that form a cycle of state 1 → 2 → 3 → 4 → 1. Since the cycle replaces one ATP molecule with an ADP and a phosphate in the solution where the motor is operated, the chemical potential ($\Delta\mu$) associated with ATP, ADP and phosphate concentrations of the solution is the energy driving the 120º forward step. $F_1$-motor's backward step synthesizes[1, 2, 21] ATP by reversing the cycle against the chemical potential (i.e. state 1 → 4 → 3 → 2 → 1). The experimentally established four-state transition diagram (Fig. 1) is a basis of the present analysis.

**Directional fidelity and associated thermodynamics.**

The conceptual basis of this study is a quantity called directional fidelity or alternatively directionality[18, 19] that was introduced to measure the fidelity of a nanomotor's self-propelled motion towards the forward direction. Specifically, directionality is defined as the net probability for a forward displacement: $D = (p_f - p_b)/(p_f + p_b + p_0)$ with $p_f$, $p_b$, $p_0$ being probability for forward, backward and null displacement. The direction is perfect with $D = 1$, and is lost entirely for $D = 0$. $D$ overlaps with the forward-to-backward stepping ratio[19, 24-29] ($R$) and the ratio[30] $(p_f - p_b)/(p_f + p_b)$ used in previous theoretical studies of molecular motors. In particular, the stepping ratio is experimentally measurable, and its relation with directional fidelity will be elaborated later in this paper.

In $F_1$-motor's sustained rotation by rounds of ATP consumption (i.e. steady-state operation), the motor's transitions must form repeatable cycles. In addition to the abovementioned forward and backward rotational cycles (marked by cycle fluxes $J_c^+$ and $J_c^-$), the transitions 4→1→4 and the transitions 2→3→2 form cycles too ($J^0_{c\alpha}$ and $J^0_{c\beta}$), which but produce null displacement. Cycles may form along the other two transition pathways between states 1 and 2 and between states 3 and 4, but these cycles do not affect rotational displacements. The motor's four-state diagram has four transition pathways; but only the two linking states 4 and 1 and linking states 2 and 3 are coupled to displacements of the rotor subunit. By this feature, the four cycles $J_c^+$, $J_c^-$, $J^0_{c\alpha}$ and $J^0_{c\beta}$ sufficiently accounts for any rotational displacements. Quantitatively, these cycle fluxes combine to yield transition fluxes for all the four transitions along the two displacement-coupled pathways: $J_{4,1} = J_c^+ + J^0_{c\alpha}$, $J_{1,4} = J_c^- + J^0_{c\alpha}$, $J_{2,3} = J_c^+ + J^0_{c\beta}$, and $J_{3,2} = J_c^- + J^0_{c\beta}$. Here $J_{4,1} = p_4 k_{4,1}$ is the flux accompanying the 4→1 transition with $p_4$ being the occupation probability for state 4; other three transition fluxes are similarly defined. Therefore, the cycle fluxes $J_c^+$, $J_c^-$, and $J^0_{c\alpha} + J^0_{c\beta}$ are identified as the displacement probabilities $p_f$, $p_b$ and $p_0$, respectively. This yields the directionality for $F_1$-motor as $D = (J_c^+ - J_c^-)/(J_c^+ + J_c^- + J_{c\alpha}^0 + J_{c\beta}^0)$.

$F_1$-motor's sustained directional motion is characterized by a steady-state directionality, which is in turn decided by the steady-state fluxes for the four elemental cycles (i.e. $J_c^+$, $J_c^-$, $J^0_{c\alpha}$ and $J^0_{c\beta}$). The motor's $D$ is further expressed in terms of transition fluxes as $D = (J_{2,3} - J_{3,2})/(J_{2,3} + J_{1,4})$ using the relations between the cycle fluxes and transition fluxes. Applying steady-state condition $J_{4,1} + J_{3,2} = J_{2,3} + J_{1,4}$ yields $D = (J_{4,1}/J_{1,4} - 1)(J_{2,3}/J_{3,2} - 1)/(J_{4,1}J_{2,3}/J_{1,4}J_{3,2} - 1)$. Note that the total entropy productions[31-33] of the object plus environment due to the net forward flux through the two pathways are $\Delta S_{4,1} = k_B \ln(J_{4,1}/J_{1,4})$ and $\Delta S_{2,3} = k_B \ln(J_{2,3}/J_{3,2})$ ($k_B$ is Boltzmann constant). Replacing the flux ratios in $D$ with the entropy productions yields the motor's directionality in terms of entropy productions along the two transition pathways coupled to the 40° and 80° substeps. Namely,

$$D = \frac{(e^{\Delta S_{2,3}/k_B} - 1)(e^{\Delta S_{4,1}/k_B} - 1)}{e^{(\Delta S_{2,3} + \Delta S_{4,1})/k_B} - 1}.$$  (1)

This directionality-entropy relation had been obtained in ref.[19] by a general consideration; here we derive it again from $F_1$-motor's transition diagram specifically. Some details of the derivation will be needed for later discussions.

**Maximum directional fidelity.**

The fuel energy is either converted to work against an opposing torque or dissipated into heat, namely $\Delta\mu = T(\Delta S_{1,2} + \Delta S_{2,3} + \Delta S_{3,4} + \Delta S_{4,1}) + F \cdot d$ by energy conservation. Here $T$ is the temperature, $F$ is torque applied to the motor, and $d$ is the step size (120°). The fuel energy is consumed by $F_1$-motor along the four transition pathways, i.e., $\Delta\mu = \Delta\mu_D + \Delta\mu_v$ with $\Delta\mu_D$ and $\Delta\mu_v$ as the energy consumption processed by the two $D$-relevant transition pathways (i.e., state 2 $\leftrightarrow$ state 3 and state 4 $\leftrightarrow$ state 1) and the other two pathways. As the $D$-relevant pathways are the only channels by which rotational displacements occur and work may be done, the work competes with the dissipation $T(\Delta S_{2,3} + \Delta S_{4,1})$ from the same source of energy consumption $\Delta\mu_D$. Namely $\Delta\mu_D = T(\Delta S_{2,3} + \Delta S_{4,1}) + F \cdot d$ and $\Delta\mu_v = T(\Delta S_{1,2} + \Delta S_{3,4})$.

Applying Lagrange multipliers to eq. 1 under the energy constraint $\Delta\mu_D = T(\Delta S_{2,3} + \Delta S_{4,1}) + F \cdot d$ yields the highest possible $D$ for a certain value of $\Delta\mu_D$ and $F$ as $D_{\max}(F) = \tanh[(\Delta\mu_D - F \cdot d)/4k_B T]$. Namely,

$$D_{\max}(F) = \frac{e^{(\Delta\mu_D - Fd)/2k_B T} - 1}{e^{(\Delta\mu_D - Fd)/2k_B T} + 1}.$$  (2)

The thermodynamic condition for the optimal $D$ is $\Delta S_{2,3} = \Delta S_{4,1} = (\Delta\mu_D - F \cdot d,)/2T$.

**Maximum speed on top of fidelity optimality.**

$F_1$-motor's average speed $v$ is readily obtained from the steady-state solution to the four-state transition diagram. Specifically, $v$ is given by the net forward transition flux along any of the four pathways, which is equal to one another for the steady state of a branchless circular chain of transitions like $F_1$-motor's. Namely $v/d = p_i k_{i,i+1} - p_{i+1} k_{i+1,i} = p_i k_{i,i+1}[1 - exp(-\Delta S_{i,i+1}/k_B)]$ with the transition from state $i$ to state $i+1$ along the motor's forward rotational direction (indicated by the dashed circle in Fig. 1). Applying $\sum p_i = 1$ yields $(v/d)\sum 1/k_{i,i+1}[1 - exp(-\Delta S_{i,i+1}/k_B)] = 1$. Namely

$$\frac{d}{v} = \frac{1}{k_{1,2}(1-e^{-\Delta S_{1,2}/k_B})} + \frac{1}{k_{2,3}(1-e^{-\Delta S_{2,3}/k_B})} + \frac{1}{k_{3,4}(1-e^{-\Delta S_{3,4}/k_B})} + \frac{1}{k_{4,1}(1-e^{-\Delta S_{4,1}/k_B})}. \quad (3)$$

Eq. 3 ensures zero speed at the stall torque when the work done by $F_1$-motor exhausts $\Delta\mu_D$ to nullify the $D$-associated dissipation, i.e., $T(\Delta S_{1,2} + \Delta S_{3,4}) \to 0$. Eq. 3 also exposes an extra limit to the motor's speed: A finite dissipation must accompany the two pathways uncoupled from directionality, i.e. $\Delta\mu_v = T(\Delta S_{1,2} + \Delta S_{3,4}) > 0$; otherwise the two pathways would be perfectly reversible and infinite slow to render a zero speed even at zero load. This part of energy dissipation becomes the primary limit to $F_1$-motor's speed particularly at zero-load operation as the $D$-associated dissipation is high with vanishing work.

The equal net forward flux along each transition pathway leads to an iterative entropy-rate relation $k_{i,i+1}[1 - exp(-\Delta S_{i,i+1}/k_B)] = k_{i,i-1}[1 - exp(-\Delta S_{i,i-1}/k_B)]$ for $F_1$-motor's branchless transition diagram. Applying the entropy-rate relation to eq. 3 yields the speed in a new formula that only depends on the entropy productions and rates of the two pathways pertinent to the extra speed limit.

$$v = \frac{d \cdot (e^{\Delta S_{1,2}/k_B} - 1)(e^{\Delta S_{3,4}/k_B} - 1)}{(e^{\Delta S_{3,4}/k_B} - 1)(e^{\Delta S_{1,2}/k_B}/k_{1,2} + 1/k_{2,1}) + (e^{\Delta S_{1,2}/k_B} - 1)(e^{\Delta S_{3,4}/k_B}/k_{3,4} + 1/k_{4,3})}. \tag{4}$$

On top of the $D$ optimization, the highest possible speed is obtained by maximizing the speed using the entropy productions $\Delta S_{1,2}$ and $\Delta S_{3,4}$ as variables under the constraint $\Delta \mu_v = T(\Delta S_{1,2} + \Delta S_{3,4})$. The highest speed for a certain dissipation $\Delta \mu_v$ is

$$v_{0,\max} = \frac{v_{\max}}{k_c d} = \frac{e^{\Delta \mu_v/k_B T} - 1}{e^{\Delta \mu_v/k_B T} + 2\sqrt{ab}e^{\Delta \mu_v/2k_B T} + a + b - 1}. \tag{5}$$

Here $k_c$ is the joint rate for the ATP binding and hydrolysis: $1/k_c = 1/k_{3,4} + 1/k_{1,2}$; $a$ and $b$ are two rate ratios: $a = k_c(1/k_{3,4} + 1/k_{4,3})$ and $b = k_c(1/k_{1,2} + 1/k_{2,1})$. In eq. 5, the speed is normalized to the joint rate ($k_c$) and the step size ($d$) so that the resultant normalized speed ($v_{0,\max}$) is a dimensionless quantity with values between 0 and 1.

The four-state transition diagram of $F_1$-motor is completely specified by eight transition rates along the four pathways. Physical quantities associated with the motor's steady-state operation, such as occupation probabilities for each state ($p_i$), transition fluxes ($J_{i,j}$), cycle fluxes ($J_c^+$, $J_c^-$, $J^0_{c\alpha}$, $J^0_{c\beta}$), entropy productions ($\Delta S_{i,j}$), directionality ($D$) and speed ($v$), are all functions of the eight rate parameters. For a certain energy partition ($\Delta \mu_D$, $\Delta \mu_v$) and load ($F$), the eight parameters are reduced to six by the directionality optimization, and further reduced to four by the speed optimization. The optimal speed of eq. 5 depends explicitly on three unknown parameters ($k_c$, $a$, $b$) that are obtained by re-grouping the remaining four rates.

**Relation between directional fidelity and measured stepping ratio.**

Single-motor mechanical experiments have been used to measure a forward-to-backward stepping ratio for translational motor proteins[34, 35] and $F_1$-rotor[17]. These mechanical measurements detect a motor's major displacements, which are not necessarily the full cycles ($J_{c+}$ or $J_{c-}$) coupled with the complete fuel turnover. In the $F_1$ experiments[17] to which the present study will compare, the 40º and 80º substeps were contained in the detected displacements but not resolved. Therefore, the detected displacements correspond to the average fluxes along both displacement-coupled pathways: $J_+ = (J_{2,3} + J_{4,1})/2$ for the detected forward displacements and $J_- = (J_{3,2} + J_{1,4})/2$ for the backward ones. Hence the so-called forward-to-backward stepping ratio, obtained as the ratio of detected forward displacements over the backward ones, is $R = J_+/J_-$ instead of $R = J_{c+}/J_{c-}$. Following the relations between transition fluxes and cycle fluxes, we have $J_+ = J_{c+} + J_{0/2}$ and $J_- = J_{c-} + J_{0/2}$, with $J_{0/2} = (J_{c0,\alpha} + J_{c0,\beta})/2$. Thus the detected forward or backward displacements include both full directional steps ($J_{c+}$ or $J_{c-}$) and futile steps ($J_{c0,\alpha} + J_{c0,\beta}$), in consistence with the experimental finding[17]. The futile steps accompanying forward rotation are $k_{4,1} \rightarrow k_{1,4}$ for $J_{c0,\alpha}$ and $k_{2,3} \rightarrow k_{3,2}$ for $J_{c0,\beta}$; the futile steps accompanying backward rotation are $k_{1,4} \rightarrow k_{4,1}$ for $J_{c0,\alpha}$ and $k_{3,2} \rightarrow k_{2,3}$ for $J_{c0,\beta}$. Therefore the futile steps accompanying the full forward steps ($J_{c+}$) and the full backward steps ($J_{c-}$) are reverse of each other and equally share the fluxes of futile cycles $J_{c0,\alpha} + J_{c0,\beta}$. Directionality then can be re-constructed from $J_+$ and $J_-$ as $D = (J_+ - J_-)/(J_+ + J_-)$. Hence a fidelity-stepping ratio relation as

$$D = (R - 1)/(R + 1). \tag{6}$$

For nanomotors in general, a $D$-relevant transition pathway is not necessarily coupled to a detectable major substep. On example is a translational biomotor called kinesin: its detected single-sized displacements[34, 35] contain no substep, and are often attributed to a single transition pathway by theoretical studies[19, 24-29]. Accordingly, the measured stepping ratio is interpreted using one pathway, i.e. $R = J_{2,3}/J_{3,2}$ or $R = J_{4,1}/J_{1,4}$. This interpretation is equivalent with the $R = J_+/J_-$ interpretation for a motor of optimal fidelity (i.e. satisfying eq. 2): the condition $\Delta S_{2,3} = \Delta S_{4,1}$ leads necessarily to $J_{2,3}/J_{3,2} = J_{4,1}/J_{1,4}$ and thereby $J_{c0,\alpha} = J_{c0,\beta} = J_{0/2}$. Hence $R = J_+/J_- = J_{2,3}/J_{3,2} = J_{4,1}/J_{1,4}$. The $D$-$R$ relation of eq. 6 is equally valid for the different interpretations.

## RESULTS AND DISCUSSION

**$F_1$-motor's directional fidelity is maximized by nearly exhausting the fuel energy and for arbitrary load up to stalemate.**

A nanomotor following eq. 2 attains the highest possible $D$ for arbitrary $F$ up to the stall load. Such a unique working regime, termed universal optimality[19], exists in translational nanomotor kinesin as a theory-experiment comparison suggests. It was also suggested[19] that the regime may be accessible by adjusting a single parameter associated with a motor's physical construction. The stepping ratio for a motor of universally optimal fidelity (eq. 2) is a single exponential decay with increasing $F$. Namely, $R_{max}(F) = \exp[(\Delta\mu_D-Fd)/2k_BT] = R_0\exp(-F\delta/k_BT)$ with a characteristic coupling length $\delta = d/2$ and a load-independent pre-exponential factor $R_0 = \exp(\Delta\mu_D/2k_BT)$.

Universally optimal directional fidelity is achieved by $F_1$-motor. The $R(F)$ data collected by Toyabe et al.[17] for a fixed $\Delta\mu$ value are well fitted by a single exponential decay[17] with $\delta = 55°$, which is nearly half of the 120° step size (Fig.2).

The fit to the $R$ data yields a pre-exponential factor of $R_0 \sim 2618$, indicating a fidelity as high as $\sim 99.92\%$ for forward rotation per fuel consumption under zero torque. The $R_0$ value yields the amount of energy used by $F_1$-motor to produce directionality: $\Delta\mu_D = 15.74$ $k_BT$, which almost equals the entire chemical energy released from ATP hydrolysis, i.e., $\Delta\mu = 15.89 \pm 0.75$ $k_BT$ as measured in the experiment[17]. Hence the motor uses its fuel energy exhaustively to produce the highest possible $D$ near the limit of energy conservation.

**$F_1$-motor's extreme fidelity maximization encompasses its extreme energy conversion.**

Since the stall torque nullifies directionality or levels the stepping ratio (i.e., $D(F_s) = 0$ or $R(F_s) = 1$) to bring $F_1$-motor to stalemate, this motor necessarily has an extreme stall torque $F_s = \Delta\mu_D/d \approx \Delta\mu/d$ following eq. 2. Hence $F_1$-motor's extreme directionality of universal optimality adequately leads to extreme energy conversion (i.e., $F_s \times d \approx \Delta\mu$) at the stall torque. The extreme energy conversion is also possible[19] outside the regime of universal optimality, if a motor's $D$ is optimized merely for the stall load. But this motor's $D$ at lower load is normally lower[19] than $D_{max}(F)$ of eq. 2.

**$F_1$-motor's finite-speed operation and energy dissipation.**

Toyabe et al.[17] measured the stall torque for a wide range of $\Delta\mu$ and ATP concentrations, and always found $F_s \times d \approx \Delta\mu$ within the experimental error for $\Delta\mu$ determination ($\sim \pm 0.75$ $k_BT$). Since $\Delta\mu_D = F_s \times d$ holds for $F_1$-motor, this motor must use an extremely small dissipation – within the experimental error of $\Delta\mu$, i.e. $\Delta\mu_v \leq 0.75$ $k_BT$ – to support its finite-speed operation. This part of energy dissipation ($\Delta\mu_v$) is the primary limit to $F_1$-motor's speed at low loads as previously discussed in Methods. An immediate question is: is such a tiny dissipation enough to account for the measured speed of the motor? To answer this question, we quantify the best possible speed and confront it with the $F_1$ experiments.

The highest possible speed ($v_{0,max}$ of eq. 5) for a fixed $\Delta\mu_v$ value exhibits a symmetric stingray-like pattern, which has two peaks when the two rate ratios ($a$, $b$) take certain values: $a \to 1$, $b \to 0$ for one peak and $a \to 0$, $b \to 1$ for the other (Fig. 3A). The first peak is generally accessible for any motor when the fuel concentration approaches zero: the fuel binding rate invariably approaches zero at this limit so that $1/k_{3,4} \to \infty$, $a \to 1$ and $b \to 0$ regardless of a motor's enzymatic capacity. However, accessibility of the other peak of the stingray map depends on the enzymatic capacity of a motor.

Toyabe et al.[17] measured speed of $F_1$-motor at various ATP concentrations but for a fixed $\Delta\mu$ value; the identical $F_s$ found for all the concentrations suggests a constant $\Delta\mu_v$ no more than 0.75 $k_B T$ as discussed above. Hence the speed data are comparable to eq. 5. But the comparison is practically possible only for zero torque for which the rates involved in eq. 5 are available for $F_1$-motor. When ATP concentration is changed, the ATP binding rate changes proportionally as $k_{3,4} = k_0 \times$[ATP] so that $a$ and $b$ vary simultaneously along a straight line of $a = \gamma b + 1$ in the $a$-$b$ plot. The $\gamma$ factor is concentration-independent as $\gamma = (1/k_{4,3} - 1/k_{1,2})/(1/k_{1,2} + 1/k_{2,1})$. The zero-torque rates of $F_1$-motor have two typical patterns[36]: $k_{1,2} \approx k_{2,1}$ (i.e., similar rates for ATP hydrolysis and resynthesis) and $k_{1,2}/k_{4,3} \approx 0.007 << 1$ (fast ATP dissociation). Hence $\gamma \approx - 0.5 \times (1 - k_{1,2}/k_{4,3}) \approx - 0.5$. Varying ATP concentration changes $a$, $b$ along the line $a = -0.5 \times b - 1$ in the $a$-$b$ plot, carving the stingray map into a $v_{0,max}$ ($a$, $b$) trajectory that is the highest possible speed for $F_1$-motor (Fig. 3B). Fig. 3C compares the $v_{0,max}$ and the zero-torque speed data normalized into $v_0 = v/dk_c$ using $k_c = k_{1,2}/(1 + k_{1,2}/k_0$[ATP]$) \approx 1640$ s$^{-1}/(1 + 3.5\mu$M/[ATP]). This $k_c$-ATP relation is deduced from reported zero-torque rates: $k_{1,2}/k_0 \approx 3.5$ μM from ref.[36], and $k_{1,2} \approx 1640$/s from refs.[20, 21, 37] for *Bacillus* PS3 $F_1$ molecules used by Toyabe et al.. The comparison shows that the optimal speed afforded by the dissipation of $\Delta\mu_v = 0.75\ k_BT$ is above the measured speed.

Toyabe et al.[17] also measured $F_1$-motor's speed for a fixed ATP concentration but different $\Delta\mu$ values. The measured maximum work ($F_s \times d$) follows $\Delta\mu$ proportionally, implying a fixed percentage of dissipation ($\Delta\mu_v/\Delta\mu$). This percentage is ~ 4% for $\Delta\mu_v = 0.75\ k_BT$. Again, the $v_{0,max}$ predicted using the 4% dissipation is above the zero-torque speed data normalized to $v_0 = v/dk_c$ (Fig. 3D).

The speed data from both ATP-varying and $\Delta\mu$-varying measurement therefore can be sufficiently explained by the small dissipation of $\Delta\mu_v = 0.75\ k_BT$, if the motor adopts the optimal speed. The conclusion remains even considering a large uncertainty of ~ 80% for hydrolysis rate ($k_{1,2}$). For the $k_{1,2}$ value used (~ 1640/s as from refs.[20, 21, 37]), a lower dissipation of $\Delta\mu_v = 0.12\ k_BT$ and the corresponding percentage ($\Delta\mu_v/\Delta\mu$ ~ 0.6%) actually fit both sets of speed data better in magnitude (Fig. 3C, D).

**$F_1$-motor's enzymatic rates are near the optimal pattern**

$F_1$-motor's enzymatic rates tend to keep the motor's operation near the two peaks of the stingray map. At the limit of low fuel concentrations ([ATP] → 0), the motor operates near one peak of the stingray map ($a = 0$, $b = 1$). At the limit of high concentrations ([ATP] → ∞), $1/k_{3,4} \to 0$, then $a \to k_{1,2}/k_{4,3} \approx 0.007$ and $b \to 1 + k_{1,2}/k_{2,1} \approx 2$, which is near the second peak ($a = 0$, $b = 1$) of the stingray map. When the fuel concentration fluctuates between the two limits, the motor's operation shifts along the straight line $a = \gamma b + 1$ in the $a$-$b$ plot, and is trapped around the two peaks within $0 \leq a \leq 1$, $0 \leq b \leq -1/\gamma$ if the $\gamma$ factor is negative. The positive rate ratio $b$ is capped by $b_{max} = -1/\gamma$ for $0 \leq a \leq 1$ (Fig. 4, inset) because $0 \geq \gamma b = a - 1$

≥ -1, and then $0 \leq b \leq -1/\gamma$. The trapping is most ideal with $\gamma = -1$, and vanishes with $\gamma = 0$. For values relevant to $F_1$-motor ($\gamma \approx -0.5$, $\Delta\mu_v \approx 0.75\ k_BT$), the highest speed $v_{0,max}$ varies within a narrow range of 0.24 - 0.53 when the fuel concentration changes from zero to infinite (Fig. 4).

**$F_1$-motor is likely near the highest possible speed on top of its fidelity maximization.**

Three aspects of the present theory-experiment comparison suggest F1-motor's speed optimality on top of fidelity maximization. First, this fidelity-speed optimization rationalizes the magnitude of the speed data using a $\Delta\mu_v$ dissipation within the experimental error of the fuel energy determination (less than 1 $k_BT$). Second, the dual optimization also captures, to some extent, the trends of the speed data from both ATP-varying and $\Delta\mu$-varying measurement. The data ($v_0$) rise slowly with $\Delta\mu$ changing from 60 to 90 pN nm, but remain rather flat with the ATP concentration changing from 0.08 μM to 250 μM. The first trend is reproduced quantitatively. The second trend is not reproduced as well quantitatively, but the data and theory both vary within a factor of two over the measured concentrations, hence are comparable in their overall flat pattern. Third, the fidelity-speed optimization predicts a stingray-like pattern in which the path linking the two peaks corresponds to the ultimately highest possible speed. The enzymatic rates of $F_1$-motor are near this unique best-speed path.

**The directional fidelity-stepping ratio relation revisited**

$F_1$-motor's stepping ratio data rule out the interpretation of $R = J_{c+}/J_{c-}$ as the name "stepping ratio" tends to imply[17]. The energy conservation $\Delta\mu = T(\Delta S_{1,2} + \Delta S_{2,3} + \Delta S_{3,4} + \Delta S_{4,1}) + Fd$ is related to the transition rates as[32] $\Delta\mu - Fd = k_BT\ln(k_{1,2}k_{2,3}k_{3,4}k_{4,1}/k_{1,4}k_{4,3}k_{3,2}k_{2,1})$ following the definitions of entropy productions. The rate ratio behind the logarithm sign is nothing but the ratio between the forward cycle flux ($J_c^+$) versus the backward cycle flux ($J_c^-$). Hence $J_{c+}/J_{c-} = \exp[(\Delta\mu - Fd)/k_BT]$. This relation can also be derived[26, 38] from fluctuation theorems or detailed balance and found applicable to single motors. For the experiments of Toyabe et al.[17], the chemical potential difference $\Delta\mu$ is a constant, and this amount of chemical energy is nearly all used for $F_1$-motor's forward motion and should factor into the $J_{c+}/J_{c-}$ ratio. By the interpretation of $R = J_{c+}/J_{c-}$, the measured stepping ratio $R(F) = R_0\exp(-F\delta/k_BT)$ must have a full-step coupling length $\delta = d$ and a load-independent pre-exponential factor $R_0 = \exp(\Delta G/k_BT)$ with $\Delta G = \Delta\mu$. The data follow a single exponential form but deviates in both $\delta$ and $\Delta G$ by a factor of 1/2, excluding this $R$ interpretation. Instead, the $\delta$ parameter being merely half of the full step size is a signature of $F_1$-motor's universal optimality of directional fidelity. This is true for translational motor kinesin too as found by a previous study[19]. But the $R = J_{c+}/J_{c-}$ interpretation might not be as conclusively excluded by the previous kinesin experiments due to unknown $\Delta\mu$ in these experiments and also due to the fact that only a portion (~ 60%) of $\Delta\mu$ is used for directional stepping and should factor into the $J_{c+}/J_{c-}$ ratio.

As mentioned in Methods, previous studies[19, 24-27, 29] often associated the measured stepping ratio to only one of the two displacement-coupled pathways. Care must be taken to interpret the physical meaning of $\delta$ and $R_0$ parameters in this alternative interpretation, though it is equivalent to the interpretation of $R = J_{c+}/J_{c-}$ for a motor of universally optimized directional fidelity like $F_1$-motor and kinesin. Bier[27] had noticed a half-step $\delta$ in kinesin's $R$ data, and interpreted it from force-biased Brownian motion in this motor's power stroke that corresponds to one of the displacement-coupled pathways. Bier deduced the ½ factor in $\delta = d/2$ from the geometry of kinesin's power stroke, namely the average position of kinesin's center-of-mass lies at the middle of a full step *d* when no load is applied. And the derivation used Boltzmann distribution though it is only applicable to thermal equilibration. The present study is different from Bier's in a few aspects. First, the ½ factor in this study is due to the optimal fidelity and the presence of two pathways which are necessary to form a repeatable cycle for sustainable direction motion. Since $F_1$-motor has two uneven substeps (40° and 80°), it is unclear whether a ½ factor can still come out of a geometric analysis similar to Bier's for kinesin that has no detectable substep[34, 35]. Second, Bier interpreted the energy term in the pre-exponential $R_0 = \exp(\Delta\mu_D/2k_BT)$ as the energy involved in the pathway corresponding to power stroke; this study identifies $\Delta\mu_D$ with the energy of both displacement-coupled pathways. As Taniguchi et al. found in their experiment[39], the energy involved in kinesin's power stroke is ~ 6 $k_BT$, about half of the value ($\Delta\mu_D \sim 12\ k_BT$) required to account for the pre-exponential factor of the measured *R(F)*. Third, the distinct load dependence of *R(F)* of this study has been derived regardless of the power stroke geometry and without using the Boltzmann distribution, hence holds for both translational and rotational motors (e.g. kinesin motor and $F_1$-rotor) and for non-equilibrium steady-state operation of the motors.

While the stepping ratio from mechanical measurements is currently subject to different interpretations, the directional fidelity is a uniquely define quantity with an exclusive link to the thermodynamic quantity of entropy productions (eq. 1). Hence we choose the directional fidelity rather than stepping ratio as the major optimizing function in the present study.

**Futile steps of $F_1$-motor and its tight chemomechanical coupling**

A $\delta$ short of the full step size found from $F_1$-motor's $R(F)$ data suggests occurrence of futile steps accompanying the motor's forward and backward steps, i.e. $R = (P_f + P_0)/(P_b + P_0)$. The ratio of the futile step to the productive forward step for this motor, $R_{0/+} = P_0/P_f$ can be deduced from the $R(F)$ data using the detailed balance relation $P_f/P_b = J_{c+}/J_{c-} = \exp[(\Delta\mu - Fd)/k_BT]$. Namely, $R_{0/+} = (1 - R/\sigma)/(R - 1)$ with $\sigma = \exp[(\Delta\mu - Fd)/k_BT]$. The deduced futile-to-forward ratio increases exponentially with torque, and reaches 1 at the stall torque (Fig. 5), suggesting an equal probability for futile and forward steps (i.e. $P_0 = P_f$).

Knowing $F_1$-motor's optimal directional fidelity, i.e. $R = \sigma^{1/2}$, we can predict the futile-to-forward ratio using the same detailed balance relation. This yields $R_{0/+} = 1/\sigma^{1/2} = 1/R$, which matches the experimentally deduced $R_{0/+}$ (Fig. 5).

The high probability of futile steps near the stall torque seems to raise doubts about the well accepted notion of tight chemomechanical coupling for $F_1$-motor (namely one productive step per fuel molecule consumed), since the experiments detecting the futile steps do not rule out the possibility that they consume ATP molecules without producing any forward rotation. The $R_{0/+}$ prediction is essentially based on the assumption that all futile steps either involve dissociation and reverse binding of either ADP (i.e. $J_{0,\alpha}$) or phosphate ($J_{0,\beta}$), but no formation/breaking of chemical bonds. With the quantitative agreement between the experimentally deduced and predicted $R_{0/+}$, the detected futile steps are accounted for by the cycles/steps that neither annihilate nor create ATP. These results tend to suggest that $F_1$-motor retains a tight chemomechanical coupling up to stalemate despite detectable futile steps, a trait that likely underlies this motor's extreme performance. A tight chemomechanical coupling up to stalemate is consistent with the observation that $F_1$-motor near stalemate converts almost all of the fuel energy into work no matter how the fuel energy is changed by fuel/product concentrations. As a comparison, kinesin motor converts the fuel energy partially to work at stalemate, and its stall force follows a different dependence[35, 40] on fuel concentrations. Previous experiments[41, 42] suggest a tight chemomechanical coupling for kinesin too but for low loads. When kinesin is near stalemate under a high load, fuel-consuming futile cycles might be activated by the load as suggested by experiments[35] and theoretical analysis[28, 43], casting doubt on kinesin's tight chemomechanical coupling at stalemate.

**CONCLUSIONS**

We have applied a concept called directional fidelity to analyze recent single-motor experiments on $F_1$-motor by Toyabe et al. with controlled chemical potentials and load. This analysis exposes quantitative evidence from these and earlier experiments for the motor's multiple extreme performance.

First, $F_1$-motor evidently uses almost all available energy from the fuel to maintain the highest possible fidelity for forward motion over futile and backward motion. This extreme fidelity holds for arbitrary load up to the motor's stalemate, sufficiently resulting in an energy conversion near the limit of energy conservation. $F_1$-motor's extreme fidelity thus encompasses the motor's well-known extreme energy conversion and provides a mechanistic basis for the latter.

Second, $F_1$-motor likely attains an extreme speed on top of the directional fidelity optimality so that less than one $k_BT$ dissipation in fuel hydrolysis suffices for the motor's finite-speed high-fidelity operation. The extreme speed renders the extreme fidelity practically accessible.

Third, $F_1$-motor's enzymatic rates follow a distinct pattern that best facilitates the extreme fidelity-speed over fluctuations in fuel supply and chemical potentials. Furthermore, a tight chemomechanical coupling is probably retained by the motor up to stalemate despite presence of futile steps, as a theory-experiment comparison tends to suggest that the futile steps unlikely involve any formation/breaking of chemical bonds of the fuel molecules. The robust chemomechanical coupling and the rate pattern conducive to the extreme fidelity/speed/energy conversion imply an extreme enzymatic capability of $F_1$-motor.

Hence F1-motor's directional fidelity, speed, enzymology and energy conversion probably approach the best possible performance allowed by physical laws for all imaginable nanomotors. While this almost all-best performance is not unexpected, qualitatively speaking, for this remarkable motor, this study presents a unified framework to quantify the multiple aspects of extreme performance and their underlying physical connections. In particular, this study indicates a crucial role of directional fidelity in $F_1$-motor's superior performance, a finding that may have wide implications to high-performance nanomotors from biology as well as future nanotechnology. The unified fidelity-speed-enzyme analysis developed in this study might be extended to study other biological and artificial nanomotors.

## ACKNOWLEDGEMENT

This work is partially supported by FRC grants under R-144-000-244-133, R-144-000-259-112 (both to ZSW).

**Figure captions**

**Fig. 1 (Color online). Consensus chemomechanical cycle for $F_1$-motor's full rotational step of 120º.** The blue arrow is the rotor subunit, and the spheres are the three stator subunits that each host an enzymatic site for ATP binding and hydrolysis. The symbol T is for ATP, D for ADP and Pi for phosphate. The circle with arrow indicates the direction of a cycle of transitions (state 1→2→3→4) that consume an ATP molecule and produce a forward rotational step. The cycle is completed by transition state 4→1 that sends the motor to a state equivalent of state 1 except for a 120º counter-clockwise rotation of the rotor subunit.

**Fig. 2 (Color online). The forward-to-backward stepping ratio of $F_1$-motor.** The data (filled circles, from ref.[17]) were collected at 24.5 ºC under a controlled chemical potential ($\Delta\mu \approx 15.89 \pm 0.75 k_B T$). Different colors mark data collected from different individual $F_1$ molecules. The solid line is an exponential fit to the data.

**Fig. 3 (Color online). Speed of $F_1$-motor. A, B.** The highest possible speed predicted by eq. 5 for arbitrary enzymatic rates (**A**), and for $F_1$-motor's rates[20, 21, 36, 37] measured under zero torque (**B**). **C, D**. Speed versus ATP concentration and chemical potential. The speed data (filled circles, from ref.[17]) were collected at 24.5ºC under zero load for a controlled chemical potential (**C**) or for a fixed ATP concentration (10 μM) (**D**). The lines are predictions of eq. 5 using $F_1$-motor's rates.

**Fig. 4 (Color online). Enzymatic optimality.** Shown is the highest possible speed predicted by eq. 5 for various values of an enzymatic factor ($\gamma$). The dissipation is $\Delta\mu_v = 0.75\ k_B T$. See text for explanations.

**Fig. 5 (Color online). Futile-to-forward stepping ratio of $F_1$-motor.** The filled circles are deduced from the forward-to-backward stepping ratio data shown in Fig. 2 using a fluctuation theorem relation. The solid line is a prediction from $F_1$-motor's fidelity optimality for the experimental condition of 24.5 °C and $\Delta\mu = 15.89\ k_BT$.

# Figure 1

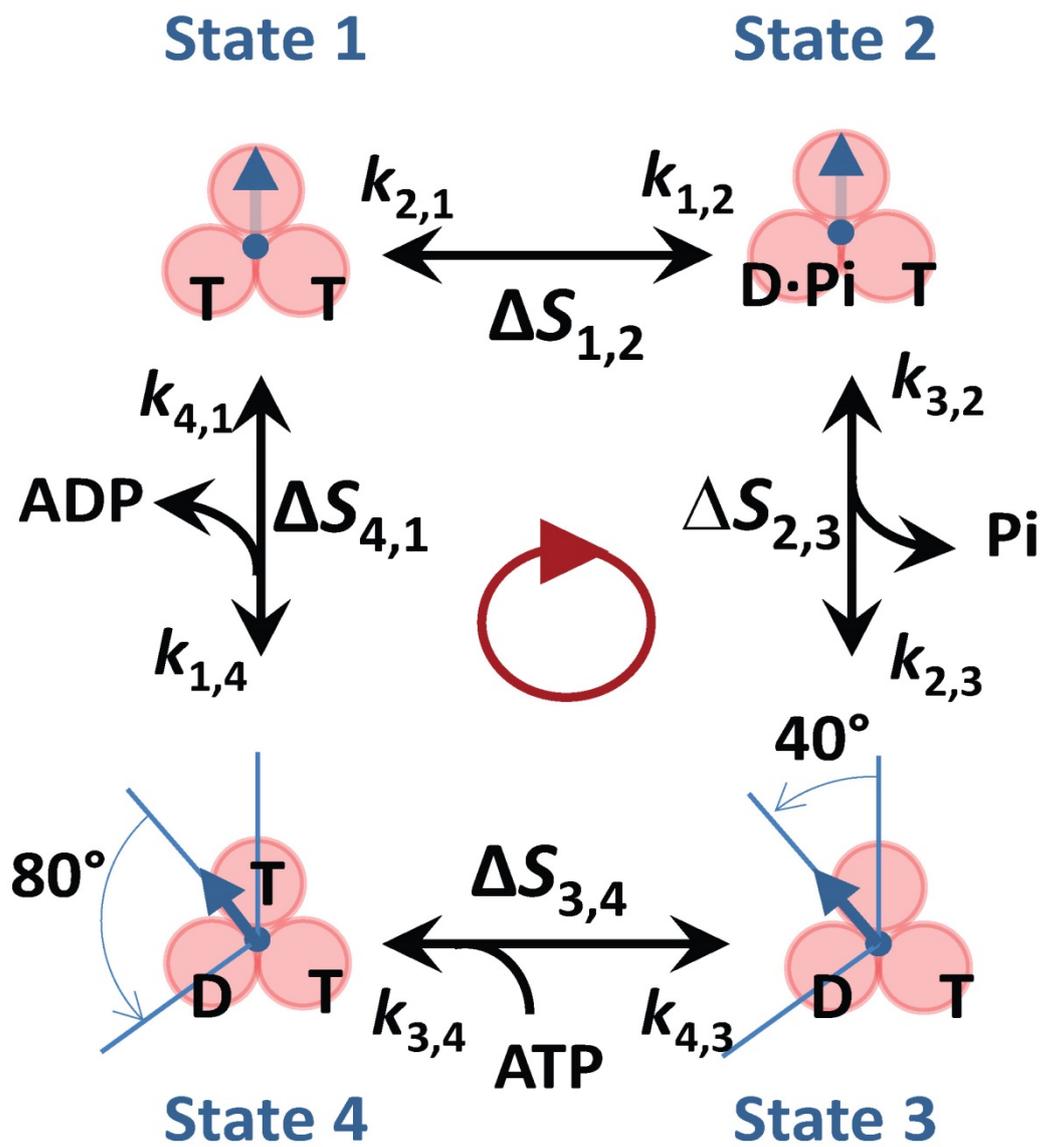

**Figure 2**

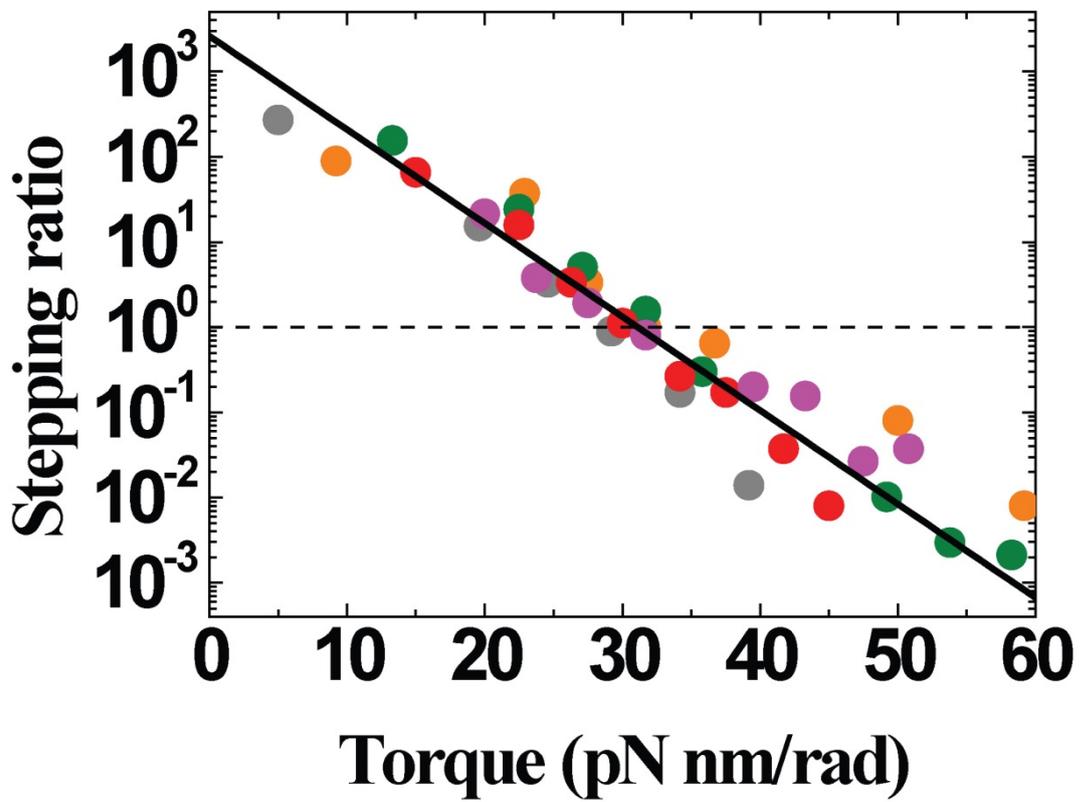

# Figure 3

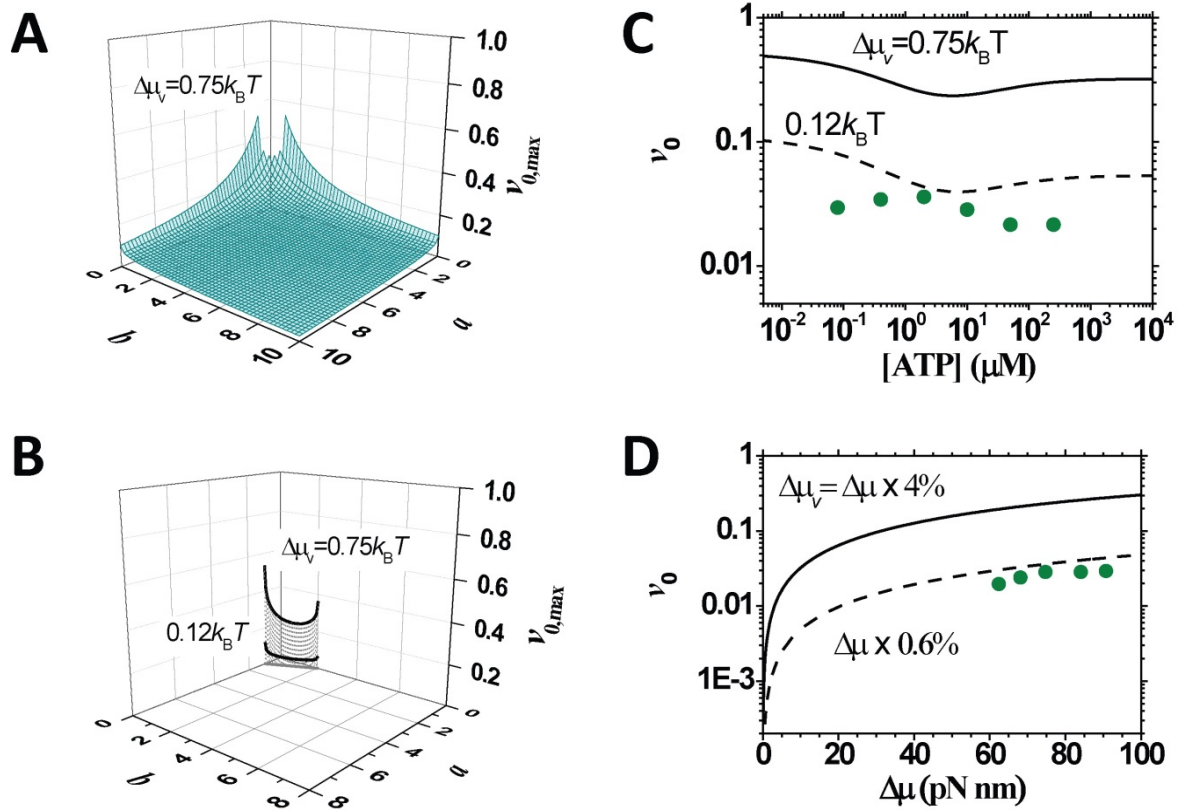

**Figure 4**

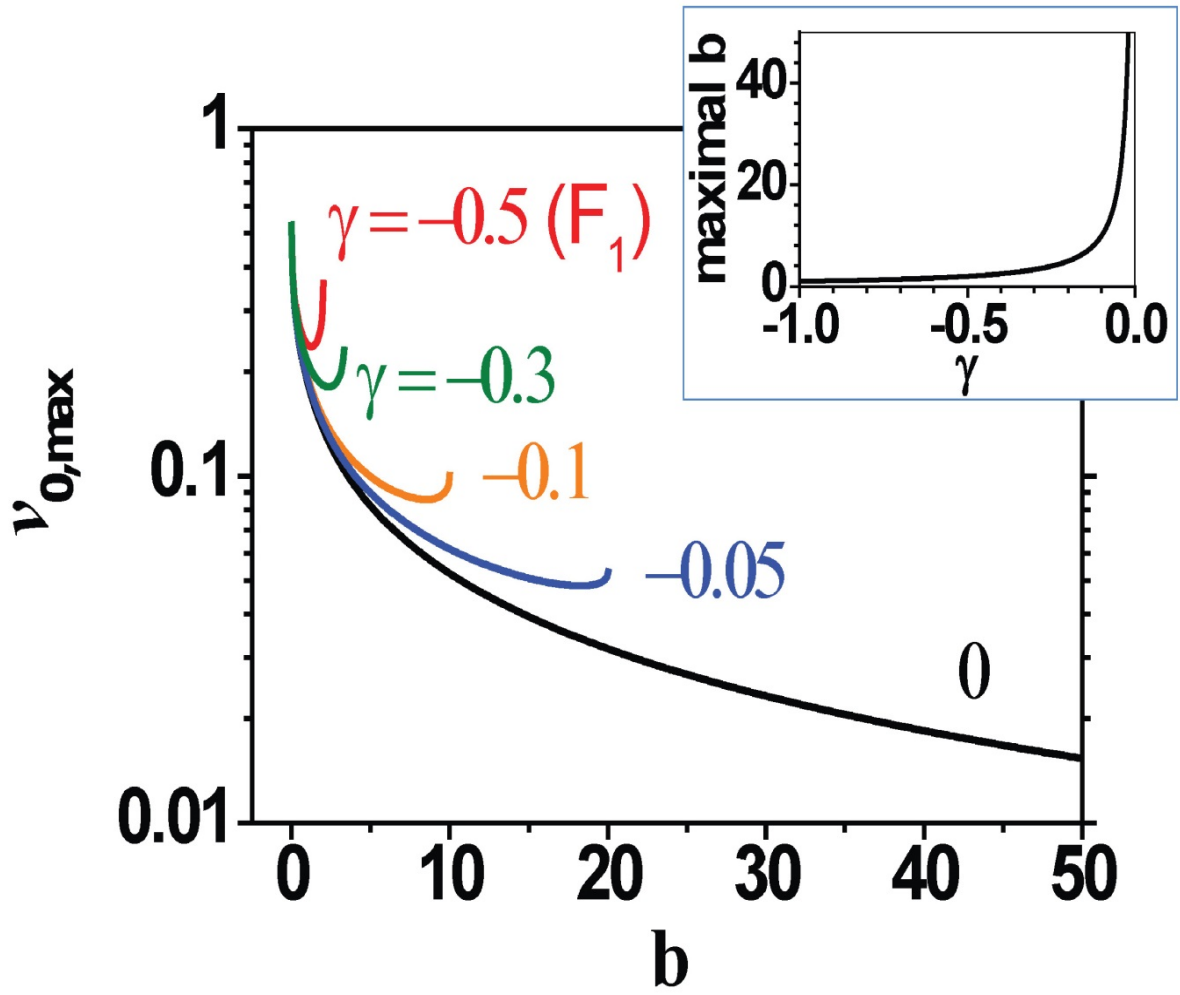

**Figure 5**

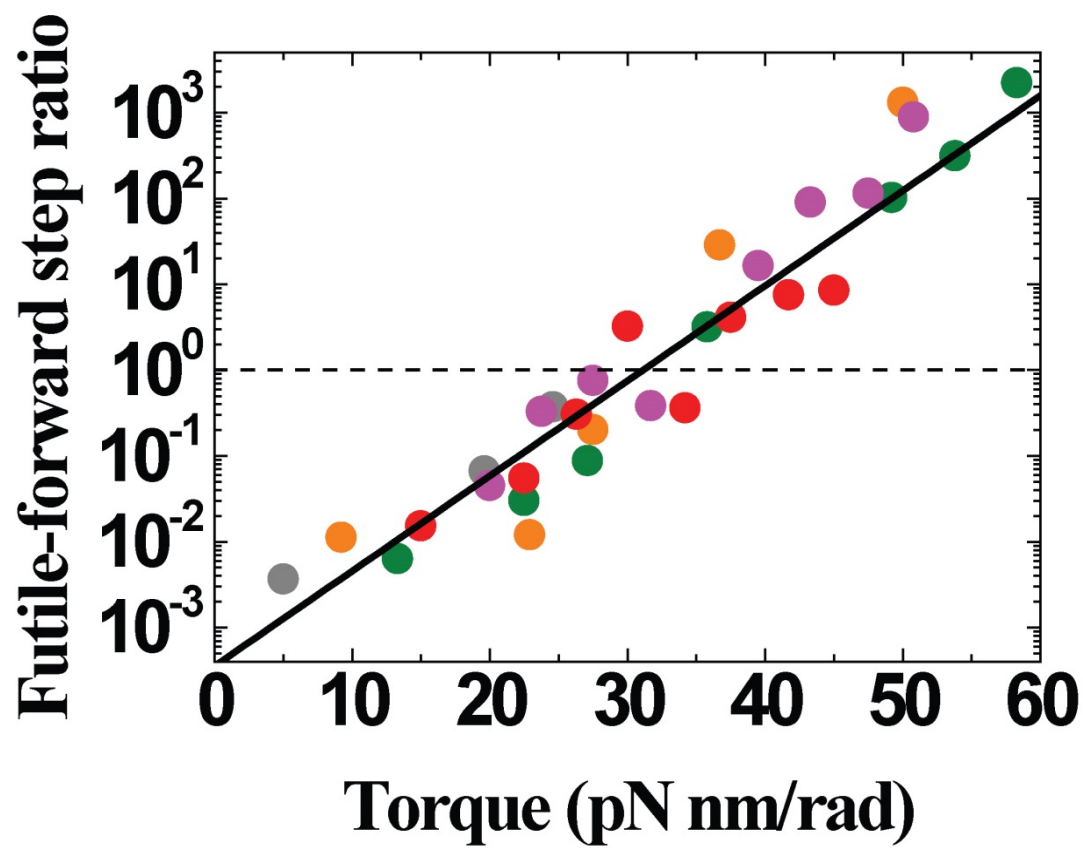